\documentclass[12pt]{article}
\usepackage{epsfig,amssymb,cite,multirow,color} 
\usepackage{amsmath}
\def\ttheta  {{\tilde{\theta}}}
\def\d {{\rm d}}

\def\d         {{\rm d}}

\DeclareMathOperator{\re}{Re}
\DeclareMathOperator{\im}{Im}

\def\sqr#1#2{{\vcenter{\vbox{\hrule height.#2pt
 \hbox{\vrule width.#2pt height#1pt \kern#1pt \vrule width.#2pt}\hrule
 height.#2pt}}}}

\def\m{\mu}
\def\g{\gamma}

\makeatletter
\@addtoreset{equation}{section}
\makeatother

\def\beq{\begin{equation}}                     %
\def\eeq{\end{equation}}                       %
\def\bea{\begin{eqnarray}}                     
\def\eea{\end{eqnarray}}   

\begin{document}

\setcounter{page}{0}
\begin{titlepage}
\titlepage
\textcolor{white}{.}

\vspace{3.5cm}
\centerline{{ \bf \Large $\mathcal{N}=2$ solutions of massive type IIA    }} 
\vskip 0.5cm
\centerline{{ \bf  \Large and their Chern-Simons duals}}
\vskip 1.5cm
\centerline{ Michela Petrini$^{a}$ and Alberto Zaffaroni$^{b}$}
\begin{center}
$^a$ LPTHE, Universit\'e Paris VI, \\
4 Place Jussieu, 
75252 Paris, France
\vskip .2cm

$^b$ Dipartimento di Fisica, Universit\`a di Milano Bicocca and INFN \\
sezione Milano-Bicocca, piazza della Scienza 3, Milano 20126, Italy
\end{center}
\vskip 1.5cm  
\begin{abstract}
We find explicit $AdS_4$ solutions of massive type IIA with $\mathcal{N}=2$             
supersymmetry obtained deforming with a Roman mass the type IIA                 
supersymmetric reduction of the M theory background $AdS_4\times M^{111}$.     
The family of solutions have $SU(3) \times SU(3)$ structure                     
and isometry $SU(3)\times U(1)^2$.                                              
They are  conjectured to be dual to three-dimensional $\mathcal{N}=2$               
Chern-Simons theories with generic Chern-Simons couplings and gauge group ranks.

\end{abstract}

\end{titlepage}

\newpage

\section{Introduction}

Recent results on 2+1 dimensional superconformal Chern-Simons theories \cite{Aharony:2008ug}
have shed new light on the $AdS_4/CFT_3$ correspondence. 
A long standing problem in establishing this latter is the identification of the 
2+1 dimensional superconformal gauge theories dual to $AdS_4$ supersymmetric backgrounds.  
In the past, attempts to find duals have focused on Yang-Mills theories flowing in the IR to superconformal fixed 
points \cite{Fabbri:1999hw, Ceresole:1999zg, Billo:2000zr, Oh:1998qi}. It seems now that supersymmetric Chern-Simons theories can do a better job. 
The  $\mathcal{N}=6$ ABJM \cite{Aharony:2008ug} model nicely incorporates all relevant features of a dual theory for the M theory 
background $AdS_4\times S^7/\mathbb{Z}_k$, including the maximally supersymmetric $\mathcal{N}=8$ case. Interestingly, for large $k$  
a better supergravity description is provided by the type IIA background $AdS_4\times \mathbb{P}^3$. A similar construction has been extended to
models with less supersymmetry. Examples of superconformal Chern-Simons theories with $\mathcal{N}=3,4,5$
supersymmetry  have been 
studied in \cite{Benna:2008zy,Imamura:2008nn,Hosomichi:2008jd,Hosomichi:2008jb,Terashima:2008ba,Hanany:2008qc,Aharony:2008gk,Jafferis:2008qz,Fuji:2008yj}.
The properties of $\mathcal{N}=2$ theories have been investigated in \cite{Martelli:2008si,Hanany:2008cd} and many models have been constructed and studied
in \cite{Hanany:2008fj,Franco:2008um,Hanany:2008gx,Davey:2009sr,Franco:2009sp,Amariti:2009rb}. For models with $\mathcal{N}=2$ supersymmetry the reduction to type IIA is still less studied.

In this paper we consider the particular case of the $\mathcal{N}=2$, M theory solution $AdS_4\times M^{111}$, its reduction to
type IIA and its supersymmetric deformations.  In particular we find a family of $\mathcal{N}=2$ supersymmetric $AdS_4$ vacua in massive 
type IIA supergravity with $SU(3)\times U(1)^2$ isometry, which include the $AdS_4\times M^{111}$ reduction as special case.

The interest in such solutions is two-fold. 
On one side, they provide non-trivial  examples of $AdS_4$ supersymmetric vacua of massive type IIA.
In spite of the  many known  $AdS_4$ vacua, the picture we have so far is not 
exhaustive.  In particular, most of the solutions have $SU(3)$ structure  
\cite{BC04,LT04,Tomasiello:2007eq,KLT08, Koerber09}, which is the simplest case
but not the generic one. Only very recently, the conditions for $\mathcal{N}=1$ supersymmetric 
solutions with generic $SU(3)\times SU(3)$ structure have been explicitly written and a type IIA solution
was given \cite{LT09}. As we will see later, in order to have non zero Roman mass and a running dilaton,
type IIA backgrounds  must have  $SU(3)\times SU(3)$ structure. In that respect, our solution is one of the first non-trivial
examples of $AdS_4$ backgrounds with $SU(3)\times SU(3)$ structure.

On the other side, the solution is also relevant for  the $AdS_4\times CFT_3$ correspondence. 
As noticed in \cite{Gaiotto:2009mv, Gaiotto:2009yz}, the Roman mass can be interpreted as the overall Chern-Simons coupling in the dual gauge
theory. More generally, all the integer Chern-Simons couplings and the ranks of the gauge groups should appear in the dual supergravity description. 
The authors of \cite{Gaiotto:2009mv, Gaiotto:2009yz} analysed the ABJM and ABJ models, finding solutions with $\mathcal{N}=0$ and $\mathcal{N}=1$ 
supersymmetry which are deformations of $AdS_4\times \mathbb{P}^3$ and have a field theoretical interpretation. 
They also find analogous solutions with 
$\mathcal{N}=2$ and $\mathcal{N}=3$ supersymmetry at first order in perturbation theory \cite{Gaiotto:2009yz}, the entire solution still remaining 
to be found. 
The same argument applies to all $\mathcal{N}=2$ Chern-Simons quivers. In particular it should always exist a massive type IIA deformation of 
the original supergravity solution which corresponds to a quiver with arbitrary Chern-Simons couplings  
and ranks. The deformation should preserve the same $\mathcal{N}=2$ supersymmetry and the same global symmetry as the original theory. 
The solution we find in this 
paper corresponds to the supergravity backgrounds  $M^{111}/\mathbb{Z}_k$.  A candidate dual Chern-Simons quiver has been proposed 
in \cite{Martelli:2008si} and further studied in \cite{Hanany:2008fj}.  It is based on a superpotential with manifest $SU(3)\times U(1)^2$ symmetry. The existence of a supergravity solution with the same symmetry can be seen as a partial check of the correctness of the proposal.

We chose $M^{111}$ because of its large global symmetry\footnote{In this case the global symmetry is  larger than 
in the corresponding $\mathcal{N}=2$ deformation of ABJM, where the full solution is still lacking. To the best of our knowledge, the type IIA reduction of $M^{111}$ enjoys the largest global symmetry among the $\mathcal{N}=2$ models with known or proposed Chern-Simons duals.}. The isometries will allow to reduce the supersymmetry conditions to a set of ordinary first order equations. Fortunately
these equations are not over-constrained and reduce to a pair of equations for two unknowns, which can be used to show the existence of a regular 
deformation. We will study the equations numerically and perturbatively.   The $AdS_4\times CFT_3$ correspondence suggests the existence of infinitely many other $\mathcal{N}=2$ supergravity solutions associated with all Sasaki-Einstein manifolds 
with dual Chern-Simons quivers\footnote{For example, $M^{111}$ belongs to the family of $Y^{p,q}(\mathbb{P}^2)$ Sasaki-Einstein
manifolds \cite{Gauntlett:2004hh,Martelli:2008rt}. These models possess the same $SU(3)$ symmetry and simply correspond to different choices of Chern-Simons couplings in the dual quiver. We shall discuss the relation of our results to $Y^{p,q}(\mathbb{P}^2)$ in the following. }. The methods of \cite{Gaiotto:2009mv} and this paper still apply. However the smaller symmetry
makes it more difficult to find explicit solutions to all orders.  For example, in the case of another famous coset manifold
$Q^{111}$, studied in the Chern-Simons context in  \cite{Franco:2008um,Franco:2009sp}, the global symmetry $SU(2)^3$ is reduced to a single  $SU(2)$. Generic Sasaki-Einstein manifolds are even more problematic having only abelian isometries. \\

The paper is organized as follows. In Section 2 we review the M theory compactification on $M^{111}$, its reduction to type IIA and 
the proposed dual
quiver. In Section 3, we study how supersymmetry is realized in type IIA. In Section 4,  we study the conditions 
for supersymmetric  massive type IIA deformations with $SU(3)\times SU(3)$ structure. We adopt an $SU(3)\times U(1)^2$ invariant ansatz, and we show that the system 
of supersymmetry 
equations is not over-constrained. We determine algebraically all the  quantities appearing in the ansatz in terms of two unknowns
for which  we write a pair of coupled first order differential equations. In Section 5, we analyze numerically and perturbatively the solution, 
showing
that it is regular, and we determine the quantization conditions on the parameters. We then interpret the result in terms of the dual Chern-Simons quiver. 
In the two Appendices, the conventions for the complex geometry of $\mathbb{P}^2$ and the supersymmetry conditions
for $SU(3)\times SU(3)$ structures are reported.

\section{The $AdS_4\times M^{111}$ background and its dual}

M theory admits $\mathcal{N}=2$ supersymmetric Freund-Rubin solutions  of the form $AdS_4\times H$  for every
Sasaki-Einstein seven-manifold $H$. In this paper we will focus on the homogeneous space
$H=M^{111}$, popular in the eighties, at the time of the Kaluza-Klein program, due to its intriguing isometry group,
$SU(3)\times SU(2)\times U(1)$. We will make use of this large isometry to find new $AdS_4$ solutions with $\mathcal{N}=2$ supersymmetry. \\

$M^{111}$ is a $U(1)$ bundle over $\mathbb{P}^2\times \mathbb{P}^1$. The metric reads \cite{Castellani:1983mf,Page:1984ad}\footnote{ In this and the following Section,
we set for simplicity the cosmological constant $\Lambda=-3|\mu|^2=12$. We will set also
$g_s=1$ for the asymptotic value of the dilaton in type IIA. These quantities can be easily reintroduced by rescaling the metric (\ref{metricaM111}) and the RR fluxes  $F_{RR}\rightarrow \frac{1}{g_s} F_{RR}$.}
\begin{equation}
\label{metricaM111}
\d s^2_{M^{111}} = \Big[ \d s^2_{\mathbb{P}^2}  +\frac{1}{8} 
( \d \theta^2+\sin^2 \theta \d \phi^2 ) + \frac{1}{64} (
\d \tau + \lambda +2\cos{\theta}
\d \phi )^2  \Big] \, .
\end{equation}
$\tau$ is an angle with period $4\pi$, while the one-form 
\beq
\lambda = - 3 \sin^2 \mu \, ( {\rm d} \psi + \cos \ttheta \, {\rm d} \tilde\phi)
\eeq
satisfies $\d \lambda =16 j_0$, where $j_0$ is the K\"ahler form on $\mathbb{P}^2$. For convenience of the reader, the metric for $\mathbb{P}^2$ 
and its natural complex structure are reported in Appendix \ref{appp1}, together with a discussion of our conventions.
$M^{111}$ can be also described as the homogeneous space 
\begin{equation}
\frac{SU(3)\times SU(2)\times U(1)}{U(1)\times U(1)} \, .
\end{equation} 
Such a characterization helped in the study of the KK spectrum. \\

Various properties of $M^{111}$,  relevant for the $AdS_4\times CFT_3$ correspondence, have been analysed a 
long time ago in \cite{Fabbri:1999hw}, where the KK spectrum and the dimension of baryonic operators were studied. In the same paper, a candidate 
dual three-dimensional Yang-Mills theory was  proposed. There are various indications now-days that a better  candidate for the dual
field theory is a  Chern-Simons theory. An $\mathcal{N}=2$ Chern-Simons theory with the right moduli space was identified in \cite{Martelli:2008si} and further studied in \cite{Hanany:2008fj}. The theory is based on a quiver with  three gauge groups and three sets of chiral fields 
$U_i,V_i,W_i$, with  $i=1,2,3$, transforming in
the $(N,\bar N,0), (0,N,\bar N)$ and $(\bar N,0, N)$ representation of the gauge groups,
respectively. They interact with the superpotential
\beq 
\mathcal{W}= \epsilon_{ijk} U_i V_j W_k \, . 
\eeq
There is a Chern-Simons coupling $k_i$ for each gauge group but no Yang-Mills terms. The theory has a global $SU(3)$ symmetry rotating the indices $i=1,2,3$ of $U,V,W$. Note that this theory has the same field content and superpotential as the quiver associated to D3-branes sitting at 
a $\mathbb{C}^3/\mathbb{Z}_3$ singularity and describing a $3+1$ $\mathcal{N}=1$ superconformal theory. 

 With the choice of Chern-Simons couplings  $k_1=k,k_2=k,k_3=-2 k$, the moduli space of the theory is the Calabi-Yau  cone $C(M^{111})/\mathbb{Z}_k$, and the theory describes the M theory background
$AdS_4\times M^{111}/\mathbb{Z}_k$. It is interesting to see in details how this happens \cite{Martelli:2008si,Hanany:2008fj}.
As discussed in \cite{Jafferis:2008qz,Martelli:2008si,Hanany:2008qc}, the D-term equations  in a $\mathcal{N}=2$ Chern-Simons theory are 
modified by a term proportional to the Chern-Simons coupling
\beq  
D_a = k_a \sigma \label{DtermI} \, .
\eeq
In this formula,  $D_a$ is the momentum map for the action of the $a$-th $U(1)$ gauge field on the elementary fields, 
and $\sigma$ is an auxiliary field in the gauge multiplets. More explicitly, the previous equations read
\bea  
\sum_{i=1}^3 |U_i|^2 - |V_i|^2 = k_1 \sigma \, ,\nonumber\\
\sum_{i=1}^3 |V_i|^2 - |W_i|^2 = k_2 \sigma \, ,\nonumber\\
\sum_{i=1}^3 |W_i|^2 - |U_i|^2 = k_3 \sigma  \, .
\label{DtermII}
\eea

These equations should be supplemented by the F-term conditions 
\beq  U_i V_j = U_j V_i \, \qquad  V_i W_j = V_j W_i \, \qquad  W_i U_j = W_j U_i \, \qquad \qquad i\ne j \, .\eeq

The sum of the three equations in (\ref{DtermII}) is zero,
reflecting the fact that the overall $U(1)$ acts trivially on the fields. The difference of the first two equations in   (\ref{DtermII}) 
imposes the vanishing of the momentum map for $U(1)_1-U(1)_2$. The last equation just determines the value of $\sigma$. 
We see that effectively we should only divide by the (complexified) gauge group  $U(1)_1-U(1)_2$\footnote{As usual, a momentum map condition and 
the modding by the corresponding
gauge group can be combined into the modding by the complexified gauge group. } which acts with charge $+2$ on $U_i$ and charge $-1$ on $V_i$ 
and $W_i$. $U(1)_3$ is broken to $\mathbb{Z}_k$ by the Chern-Simons interactions
and remains as a global symmetry.\\

For $k=1$ the theory has symmetry $SU(3)\times SU(2)\times U(1)$, where
$SU(3)$ acts on the $i,j,k$ indices, $U(1)$ is the R-symmetry, which  rotates all the
fields  $(U,V,W)$ with the same charge, and, finally, $SU(2)$ is the enhancement of $U(1)_3$ obtained by considering the doublets $R^A_i= (V_i,W_i)$. The  gauge 
invariant chiral operators are given by
\beq 
O_n=(U R R)^n \, , 
\eeq
where the $SU(3)$ and $SU(2)$ indices are symmetrized due to the F-term conditions.  Since the superpotential $\mathcal{W}$ must have R-charge two,  
$O_n$ has R-charge $2n$. In conclusion, there is exactly one chiral multiplet with R-charge $2n$ transforming in the $[3n,0]$ representation 
of $SU(3)$ and the $2n$ representation of $SU(2)$.  
We recognize the KK spectrum of M theory compactified on  $M^{111}$ \cite{Fabbri:1999hw}. A more geometrical proof, based on toric geometry, that the moduli space of the Chern-Simons theory is $C(M^{111})$ can be found in \cite{Martelli:2008si,Hanany:2008fj,Forcella:2008bb}. \\

For $k\ne 1$ we have to mod by $\mathbb{Z}_k\in U(1)_3$, which acts with charge $+1$ on $W_k$ and charge $-1$ on $V_j$. The moduli space is now $C(M^{111})/\mathbb{Z}_k$ and the $SU(2)$ symmetry is broken to $U(1)$. 
$\mathbb{Z}_k$ acts indeed on a circle in $M^{111}$ reducing its radius.
For large $k$ the compactification on $AdS_4\times M^{111}/\mathbb{Z}_k$ is effectively reduced to a type IIA compactification, as in the ABJM model.   It is easy to identify the action of $\mathbb{Z}_k$
with a shift of $\phi$ in the metric (\ref{metricaM111}).  Reducing along $\phi$ gives a supersymmetric type IIA background  
with non trivial dilaton, $F_2$ flux and $SU(3)\times U(1)^2$ symmetry. The metric is
\beq
\label{metric10}
{\rm d} s_{10}^2 = e^{2 A} \, {\rm d}s^2_{AdS_4}  + {\rm d}s^2_6 \, ,
\eeq
where the six-dimensional compact manifold is a 2-dimensional fibration over
$\mathbb{P}^2 $
\beq
\label{metric6}
{\rm d} s^2_6 =  e^{2 A} [ {\rm d}s^2_{\mathbb{P}^2} +  \frac{1}{8} \, 
{\rm d} \theta^2 + \frac{1}{32} \, \frac{\sin^2 \theta}{1 + \sin^2 \theta} ({\rm d}\tau + \lambda)^2  ] \, .
\eeq
The term $\d \tau+\lambda$ in the original metric determines a non trivial  RR two-form
\beq
\label{2form0}
F_2 =  \d \left[\frac{\cos \theta}{2(1 + \sin^2 \theta)} ({\rm d}\tau + \lambda) \right] \, .
\eeq
Finally, the warp factor is proportional to the dilaton and is given by
\beq
\label{dil}
e^{2 A} = e^{2 \varphi/3} = \frac{1}{4} \sqrt{1 + \sin^2 \theta} \, .
\eeq

Type IIA reductions
of $M^{111}$ and other Sasaki-Einstein manifolds have been considered in the past \cite{Nilsson:1984bj, Sorokin:1985ap}. The reduction was performed on the obvious $U(1)$ circle bundle with the result of breaking supersymmetry. The
natural $U(1)$ bundle of the $\mathbb{P}^2\times \mathbb{P}^1$ fibration corresponds indeed to the
R-symmetry.  A reduction along the  $\mathbb{Z}_k$ action, on the other hand, preserves supersymmetry 
since $\mathbb{Z}_k$ is a subgroup of the global $SU(2)$ symmetry. \\

In the following section we will verify that the previous solution is $\mathcal{N}=2$ supersymmetric. We will then find an $\mathcal{N}=2$ 
supersymmetric deformation of this solution in massive type IIA preserving $SU(3)$ symmetry. As suggested  in \cite{Gaiotto:2009mv} such solutions should correspond to  the case $\sum_a k_a\ne0$ and possibly generic $N_i$. 
The existence of a solution is predicted by the Chern-Simons theory since a modification of $\sum_a k_a$
and $N_i$ does not affect the superpotential and preserves the $SU(3)$ symmetry. It is quite remarkable that such
supergravity solution actually exists.

\section{Undeformed solution}



We now verify that the  background \eqref{metric10}-\eqref{dil}  is a solution of type IIA supergravity with $\mathcal{N}=2$ supersymmetry. 
$M^{111}$ admits two real Killing spinors which  give, after reducing to six dimensions,
one Weyl spinor each and, hence, $\mathcal{N}=2$ supersymmetry. We use the language of Generalised Complex Geometry \cite{Hitchin,gualtieri}
which is briefly reviewed in Appendix \ref{apppure}. \\

Since each Killing spinor can be seen as defining an $SU(3)$ structure, a convenient way 
to check supersymmetry is to look for two pairs of $SU(3)$ structure pure spinors 
satisfying \cite{gmpt1,gmpt2}
\bea
\label{susyeq1}
&& ({\rm d}-H\wedge) (e^{3A-\varphi} \im \Phi_- ) = - 3 e^{2A - \varphi} \mu \im\Phi_+  + \frac{e^{4 A}}{8} \ast \lambda(F) \, ,\\
\label{susyeq2}
&& ({\rm d}-H\wedge) (e^{2A-\varphi}  \Phi_+)= - 2  \mu  e^{A - \varphi}  \re\Phi_- \, ,
\eea
where $\mu$ is related to the cosmological constant in $AdS_4$ by $\Lambda = -3 |\mu|^2$. By
changing phases in the spinors we can always take $\mu$ real, and we will do so in the following. 

We can write the $SU(3)$ structure pure spinors as in \eqref{SU(3)ps},
choosing  the parametrization $a= i  e^{A/2} e^{i (\rho + \alpha)}$ and $x= e^{A/2} e^{i \alpha}$
\beq
\label{SU(3)pssol}
\Phi_+ = \frac{i}{8} e^{i \rho} e^A \, e^{-i J} \, ,\qquad \qquad 
\Phi_- =-  \frac{i}{8} e^{i(\rho+ 2  \alpha )} e^A \, \Omega  \, .
\eeq

The fibered structure of the six-dimensional metric suggests a natural splitting into base and fiber directions
for the choice of the holomorphic three-form $\Omega$ and K\"ahler form $J$
\beq
\label{holJf}
\Omega = i \omega \wedge z \, ,\qquad \qquad J = j + \frac{i}{2} z \wedge \bar{z} \, . \eeq 
Here $z$ is a one form on the $S^2$ fiber
\beq
z = - \frac{i e^A }{2 \sqrt{2}} 
\Big[ \d \theta + i \frac{\sin \theta}{2 \sqrt{1 + \sin^2 \theta}} ( \d \tau + \lambda ) \Big] \, ,
\eeq
while $j$ and $\omega$ are a  rotation of the natural complex structure on $\mathbb{P}^2$
(see Appendix  \ref{appp1}  for notations)
\bea
\label{su2strP2}
&& j = e^{2 A} (\cos \gamma \, j_0 + \sin \gamma  \re \hat{\omega}_0) \, ,\nonumber \\
&& \omega = e^{2 A} [ (- \sin \gamma \, j_0 + \cos \gamma \re  \hat{\omega}_0) 
+ i \im \hat{\omega}_0 ] \, ,
\eea
with  $\gamma$ a function of the angle $\theta$ on the two-sphere.

With the choice \eqref{SU(3)pssol}, the equation \eqref{susyeq2} for the even pure spinor  reduces
to the two conditions
\bea
\label{1form}
&& \d (3A - \varphi + i \rho) = 0 \, ,\\
\label{3form}
&& \d J - i H= - 2 \mu  e^{- i \rho} e^{-A }  \re(-i e^{i (\rho+2 \alpha)} \, \Omega) \, .
\eea
The first equation sets $\rho$ to a constant and implies the same
proportionality as in \eqref{dil} between the  dilaton and the warp factor
\beq
\varphi = 3 A \, .
\eeq

Since  $H=0$ in the solution, from the second equation we see that  $e^{i \rho}$ must be real.
Choosing  $\rho=0$,  it is straightforward to verify that \eqref{3form} is satisfied by the ansatz 
\eqref{holJf}-\eqref{su2strP2} with $ \alpha=  \pi/4$, $\mu=-2$  and 
\beq
\cos \gamma = \frac{\cos \theta}{\sqrt{1+\sin^2 \theta}} \, , \qquad \qquad  
\sin \gamma  = \sqrt{2} \frac{\sin \theta}{\sqrt{1+ \sin^2 \theta}}  \, .
\label{gamma}
\eeq

Similarly, equation \eqref{susyeq1} for $\Phi_-$ gives the closure of the imaginary part of $\Omega$
\beq
\d [e^{-A} \im ( \Omega)] = 0 \, ,
\eeq 
and the RR fluxes
\bea
\label{flux1}
&& F_4 = 0 \, , \qquad \qquad   e^{4 A} \ast F_6 = - 6 \, ,\\
\label{flux2}
&& F_0 = 0 \, , \qquad \qquad   e^{4 A} \ast F_2 = - \d (e^A \re \Omega) + 3  J^2 \, .
\eea
Again, it is easy to check that the ansatz \eqref{holJf}-\eqref{su2strP2} solves the equation for the closure
of $\im \Omega$ and that the $F_2$ defined in \eqref{2form0}  satisfies \eqref{flux2}.
Finally, we can take the equation for $F_6$ as a definition of the cosmological constant in the solution.\\

The discussion above proves that, reducing the $M^{111}$ background, we obtain a solution of IIA with one supersymmetry. We still
have to look for the second supersymmetry. However, it is immediate to construct a second pair of pure spinors
satisfying the equations \eqref{susyeq1},\eqref{susyeq2}. These have the same form as in 
\eqref{SU(3)pssol}, with $\Omega$ and $J$ defined as in \eqref{holJf}, but with a different 
complex structure obtained by a change of sign in the coordinates on the base
\bea
\label{su2strP2bis}
&& j = e^{2 A} (- \cos \gamma  \, j_0 + \sin \gamma   \re \hat \omega_0) \, ,\\
&& \omega = e^{2 A} [ (\sin \gamma  \, j_0 + \cos \gamma  \re  \hat \omega_0) - i \im \hat \omega_0 ] \, , \\
&& z  = - \frac{i e^A }{2 \sqrt{2}} 
\Big[ \d \theta - i \frac{\sin \theta}{2 \sqrt{1 + \sin^2 \theta}} ( \d \tau + \lambda ) \Big] \, .
\eea

As already mentioned, there are other solutions of type IIA with $\mathcal{N}=2$ supersymmetry and $SU(3)$ structure with the same global
symmetry. $M^{111}$ belongs indeed  to the larger family of Sasaki-Einstein manifolds $Y^{p,q}(\mathbb{P}^2)$
with $SU(3)\times U(1)^2$ isometry. These solutions correspond to different choices of Chern-Simons
couplings with $p = k_1+k_2, k = 2 k_1 +  k_2$ for $k_1,k_2\ge 0$ \cite{Martelli:2008rt}. The reduction to a type
 IIA background can be found and studied similarly.

\section{Deformed solution}

Now we come to massive type IIA deformations of the previous background that still preserve $\mathcal{N}=2$ supersymmetry
and have $SU(3)$ global symmetry. These are obtained by introducing the Roman mass $F_0$. 

We choose an $SU(3)$ invariant ansatz for the metric 
\beq
\label{metric10def}
{\rm d} s_{10}^2 = e^{2 A(\theta)} {\rm d}s^2_{AdS_4}  + {\rm d}s^2_6 \, ,
\eeq
where the six-dimensional compact manifold is still a 2-dimensional fibration over
$\mathbb{P}^2 $
\beq
\label{metric6def}
{\rm d} s^2_6 =  e^{2 B(\theta)} [ {\rm d}s^2_{\mathbb{P}^2} + \frac{ 1}{8} \, \epsilon^2(\theta)
{\rm d} \theta^2 + \frac{1}{64} \Gamma^2(\theta) ({\rm d}\tau + \lambda)^2  ] \, .
\eeq
It is still convenient to define a  one-form on the $S^2$ fiber
\beq
\label{zdef}
z = - i  e^{B (\theta)}  e^{-i \nu(\theta)} \Big[\frac{ \epsilon(\theta)}{2\sqrt{2}} \d \theta +  \frac{i}{8} \Gamma(\theta)( \d \tau + \lambda) \Big] \, .
\eeq
The phase $\nu$ will be fixed  shortly. Since we can redefine $\theta$, one of the functions in
the ansatz is redundant. We will use the freedom to change coordinate later.

For the fluxes, we take the natural $SU(3)$ invariant ansatz 
\bea
&& F_0 = f_0  \, , \nonumber \\
&& F_2 =   f_2(\theta)\, j_0 + \frac{i}{2} \, g_2(\theta) \,  z\wedge \bar z   \, ,\nonumber \\
&& F_4 =   f_4(\theta) \, j_0\wedge j_0 + \frac{i}{2}  \,   g_4(\theta)\, z\wedge \bar z \wedge j_0  \, ,\nonumber  \\
&& F_6 =    e^{4 B(\theta)} \frac{i}{4} \,  f_6(\theta)\,  z\wedge \bar z   \wedge  j_0^2 \, ,\nonumber\\
&& H =  h(\theta) j_0\wedge \d\theta \, .
\label{flux2def}
\eea
The ansatz is $SU(3)$ invariant since $j_0$ and $\lambda$ are.

It is easy to check
that, when  $F_0\ne 0$, the supersymmetry equations for $SU(3)$ structure pure spinors  require constant dilaton. Since the
dilaton is running even in the unperturbed solution, we are led to consider solutions with 
$SU(3)\times SU(3)$ structure. We can write the $SU(3)\times SU(3)$  structure pure spinors as in \eqref{purespSU2}, choosing  the  
dielectric ansatz \cite{Minasian:2006hv, Gaiotto:2009yz}
\bea
&& a =  i \cos\phi \, e^{i\rho} e^{i \alpha} e^{A/2}\, ,  \,\, \qquad  \qquad \quad  \,\, x = \cos\phi \, e^{i \alpha} e^{A/2} \, , \nonumber \\
&& b = -i \sin\phi \, e^{i\rho}e^{-i \alpha} e^{A/2}\, ,  \,\, \qquad   \qquad \,\, y = \sin\phi \, e^{-i \alpha} e^{A/2}\, .
\eea
Here $\rho$, $\phi$
and $\alpha$ are
functions of the angle $\theta$ on the fiber.

An $SU(3)\times SU(3)$ structure corresponds to a $4+2$ splitting on the internal metric, determined by a vector $z$.  It is natural for us to use the splitting into $\mathbb{P}^2$ and $S^2$: $z$ has been defined above and $j,\omega$   are defined as in (\ref{su2strP2}), with a possibly different function $\gamma$. This is the natural generalization of what we used for the undeformed solution. 
A quick analysis shows that we can consistently choose $\alpha=\pi/4$ as in the undeformed case. \\

We need to solve \eqref{susyeq1}, \eqref{susyeq2} for the new pure spinors. To simplify our notations,
let us notice that the 10-dimensional metric is invariant under the simultaneous rescaling of $\mu$
and $e^{A}$, so that we can reabsorb $\mu$ in the definition of $A$\footnote{Since $\Phi\sim e^{A}$, the pure spinor equations can be reformulated in terms of $e^{A}/\mu$ and $e^{3A-\varphi}$ and the phases in the dielectric ansatz. The arbitrary constant in $e^{3A-\varphi}$ can be reabsorbed by a rescaling of the RR fluxes and will be set to one in the text.}. We use this freedom to fix again $\mu=-2$.  The one-form component
of (\ref{susyeq2}), 
\beq 
{\rm d} (i e^{3 A-\varphi} \, e^{i\rho}\, \cos 2\phi  )=  - 4  e^{2 A - \varphi}  \re( i e^{i\rho} \, \sin 2\phi \,  z ) \, ,
\eeq
 immediately gives a lot of information. The fact that the right-hand side must be closed implies that it
is proportional to $\d \theta$. This can be obtained by choosing $\nu=\rho$ in \eqref{zdef}. The imaginary part of the previous equation then fixes the dilaton
\beq 
e^{3 A-\varphi} = \frac{1}{\cos 2\phi \cos \rho} \, .
\eeq
Again, for simplicity, we omitted an arbitrary constant.  As easily seen from the equations, the constant can be reintroduced by rescaling the RR fluxes, $F\rightarrow F/g_s$. 

The remaining equations give several differential and algebraic constraints for few unknown functions,
 $A,B,\Gamma,\epsilon,\rho, \phi,\gamma, f_i, \g_i$. To these constraints we have to add the Bianchi  
identities for fluxes
 \beq 
\d F - H\wedge F =0 \, .
\eeq
There are clearly more equations than unknowns. However, this  formidable system is not over-constrained and, with some patience, it can be reduced to a pair of linear differential equations for two unknowns. All other quantities can be
obtained algebraically and the Bianchi identities are automatically satisfied.\\

In order to simplify the resulting set of equations, it is convenient to use the freedom of redefining $\theta$. We can always choose coordinates where $\gamma(\theta)$ is the function defined in (\ref{gamma}). With this choice we can write a pair of linear differential equations for the quantities $\phi$ and $w=4 e^{2(B-A)}$,
\bea
\phi^\prime &=& 
\frac{ \sin 4\phi \, \cot \theta \, [w - 2 (\sin^2 2 \phi + 2 \tan^2\theta)]}
{4 [ w \, (1 + \sin^2 \theta) - 2 \cos^2 2\phi \, (2 \sin^2 \theta + \cos^2 \theta \, \sin^2 2 \phi)]} \, , \nonumber\\
w^\prime  &=& \frac {- w \, \cot\theta \, (\sin^2 2\phi + 2 \tan^2\theta) (w -4 - 4 \sin^2 2\phi)}{2 [ w \, (1 + \sin^2 \theta) - 2 \cos^2 2\phi \, (2 \sin^2 \theta + \cos^2 \theta \, \sin^2 2 \phi)]} \, .
\label{eqs1}
\eea 

All other quantities are then determined in terms of the previous ones. It turns out that $\tan \rho= -\cot\theta \sin 2\phi /\sqrt{2}$. 
The functions appearing in the metric are given by
\bea
\label{eqs2}
&& e^{4 A} =   \frac{2 \sqrt{2}}{f_0}  \frac{\sin 2 \phi \, (1+ \sin^2 \theta)}{\cos^2 2 \phi \, \sin 2 \theta} \, ,\nonumber\\
&& \epsilon = 2 \sqrt{w} \frac{\sqrt{4 \tan^2 \theta + 2 \sin^2 2 \phi}}{\sin 2 \phi \, (w - 4 \tan^2 \theta -2 \sin^2 2 \phi)} \, \phi' \, , \nonumber\\
&& \Gamma =  \frac{2}{\sqrt{w}} \, \frac{\cos \theta \, \sqrt{2 \tan^2 \theta +  \sin^2 2\phi}}{\sqrt{1 + \sin^2 \theta}} \, .
\eea
The fluxes read
\bea
&& h =  \sqrt{2}  e^{2 A}  \,  w  \, \frac{ \sin \theta \, ( \sin^2 2 \phi + 2 \tan^2 \theta)}
{\sqrt{(1 + \sin^2 \theta)} \, [ w   - ( 2 \sin^2 2 \phi + 4 \tan^2 \theta)]} \, \phi' \, ,\nonumber\\
&&f_2 = e^{-2 A}  \, \frac{\, [w  (1 + \sin^2 \theta) - 4 \cos^2 \theta \,  
(2 \tan^2 \theta +   \sin^2 2 \phi)]}{2 \sqrt{1 + \sin^2 \theta} \, \cos \theta \cos 2 \phi } \, ,\nonumber \\ 
&& g_2 = - 2 e^{-4 A} \frac{  [3 w  (1 + \sin^2 \theta) - 8 
\cos^2 \theta \,  
(2 \tan^2 \theta +   \sin^2 2 \phi)]}{(1 + \sin^2 \theta) w} \, ,\nonumber \\ 
&&  f_4=  - \, \frac{  w [w  (1 + \sin^2 \theta) - 8 \cos^2 \theta \,  
(2 \tan^2 \theta +   \sin^2 2 \phi)]  \sin 2 \phi}{ 8 \sqrt{2} \sin 2 \theta \, \cos^2 2 \phi } \, ,\nonumber \\
&&  g_4=  \frac{1}{2\sqrt{2}} e^{-2 A} \, \frac{ [3 w  (1 + \sin^2 \theta) - 4 \cos^2 \theta \, (2 \tan^2 \theta +  \sin^2 2 \phi)] \tan 2 \phi}{ \sin \theta \, \sqrt{1+\sin^2\theta} }\, ,\nonumber \\
&& f_6 = \,  \frac{3}{\sqrt{2}} f_0 \frac{ \sin 2 \theta \cos^2 2 \phi}{(1 + \sin^2 \theta) \sin 2 \phi}  .
\label{eqs3}
\eea
In all the expressions above $f_0$ is set to a constant by the Bianchi identities.\\

The solution has $\mathcal{N}=2$ supersymmetry. The second supersymmetry is obtained, as in the unperturbed case, 
by changing complex structure as in (\ref{su2strP2bis})
\bea
\label{su2strP2bisdef}
&& j = e^{2 A} (- \cos \gamma(\theta) \, j_0 + \sin \gamma(\theta)  \re \omega_0) \, ,\\
&& \omega = e^{2 A} [ (\sin \gamma(\theta) \, j_0 + \cos \gamma(\theta) \re  \omega_0) - i \im \omega_0 ] \, , \\
&& z = - i  e^{B (\theta)}  e^{-i \nu(\theta)} \Big[\frac{ \epsilon(\theta)}{2\sqrt{2}} \d \theta -  
\frac{i}{8} \Gamma(\theta)( \d \tau + \lambda) \Big] \, .
\eea
The pure spinors are given by the dielectric ansatz  with $\phi\rightarrow -\phi$, $\rho\rightarrow\rho$. 
The supersymmetry equations are then satisfied with the same metric and fluxes as before. The relations \eqref{eqs1}-\eqref{eqs3}
remain true.

\section{Analysis and interpretation of the solution}

We were not able to solve analytically the system of equations \eqref{eqs1}-\eqref{eqs3}, but we can study the properties
of the solution using perturbation theory and numerical analysis.

We have solved the equations \eqref{eqs1}-\eqref{eqs3} up to third order. The idea is to define a 
perturbative expansion where $\phi$, $\rho$ and the fluxes $H,F_4,F_0$ receive corrections 
at odd orders, while the metric and fluxes $F_2,F_6$ receive corrections at even orders. The constant $f_0$ has an 
odd expansion,  while $\mu$ an even one. At first order we find
\beq 
\phi^{(I)}= 2  c^{(I)} \cos 2\theta \, ,\eeq
with fluxes
\bea
h^{(I)} &=& \sqrt{2}  c^{(I)} \sin^3\theta  \, ,\nonumber\\
f_4^{(I)} &=& 32 \sqrt{2} \, c^{(I)} \, \frac{3\cos 2\theta -1}{\cos 2\theta -3} \, ,\nonumber\\
g_4^{(I)} &=& -64  \, c^{(I)} \, \frac{\cos 3\theta -13 \cos\theta}{(3-\cos 2\theta)^{3/2} }  \, . 
\eea
The constant $f_0$ requires a word of caution.  As one can see from equation (\ref{eqs2}), the limit $f_0\rightarrow 0$
is singular. The correct unperturbed limit is obtained by sending simultaneously $f_0\rightarrow 0$ and $\phi\rightarrow 0$. The solution of Section 3 is obtained by setting 
\beq
f_0 ^{(I)}= 64 \sqrt{2}  c^{(I)} \, .
\eeq

The metric receives corrections at second order, which can be easily determined and which depend on a second arbitrary constant. We do not report the  expressions here. We have checked the regularity of the resulting metric for all $\theta$, and, in particular, at the North and South poles  and at the equator of the sphere. Up to third order, the expansion gives a perfectly regular solution of type IIA supergravity. \\

More generally, we can study the regularity of the metric near the North and South pole by expanding
$\theta$ around $0$ and $\pi$. By solving the equations \eqref{eqs1}-\eqref{eqs3} near $\theta=0$ we find
\bea
\phi &=& \phi_1 \theta - \left (\frac{2 \phi_1}{3}+\frac{4 \phi_1^3}{3}\right )\theta^3 +O(\theta^5) \, ,\nonumber\\
w &=& w_0+ \frac{1}{2}( 4-w_0)(1+2 \phi_1^2)\, \theta^2 +O(\theta^4)   \, .
\eea
An identical expression holds at the South pole with  different parameters $\tilde\phi_1,\tilde w_0$.
The $S^2$ metric will be smooth near the poles if it reduces to the flat metric  in polar coordinates. 
It is easy to check, using equations (\ref{eqs2}), that  
\beq \epsilon\rightarrow2 \sqrt{\frac{1+2 \phi_1^2}{w_0}}\, , \qquad\qquad  \Gamma\rightarrow \sqrt{2}\epsilon (0) \theta \, ,
\eeq
at the North pole, and similarly, with $\phi_1\rightarrow \tilde\phi_1, w_0\rightarrow\tilde w_0$, at the South pole. The vector $z$ has then an expansion (fixing an arbitrary point in $\mathbb{P}^2$)
\beq z \sim  \d (\theta-\theta_P) + i (\theta-\theta_P) \frac{\d \tau}{2} \eeq
at both poles. Since  $\tau$ has period $4\pi$, this guaranties  the regularity of the metric. Warp factors
and fluxes are similarly computed and are regular.  The expansion of the equations near the equator is also smooth. \\

We can further study the solution by numerical analysis. We can use $\phi_1,w_0$ as parameters labeling the solutions of the system (\ref{eqs2}). The analysis shows that there is a two-parameter 
family of regular solutions departing from the unperturbed one. The shape of $\phi$ and $e^{2A}$  is shown in Figure 1 for special values of the parameters.  \\

\begin{figure}[h]
\begin{minipage}[t]{\linewidth}
~~~\begin{minipage}[t]{0.5\linewidth}
\vspace{0pt}
\centering
\includegraphics[scale=0.8]{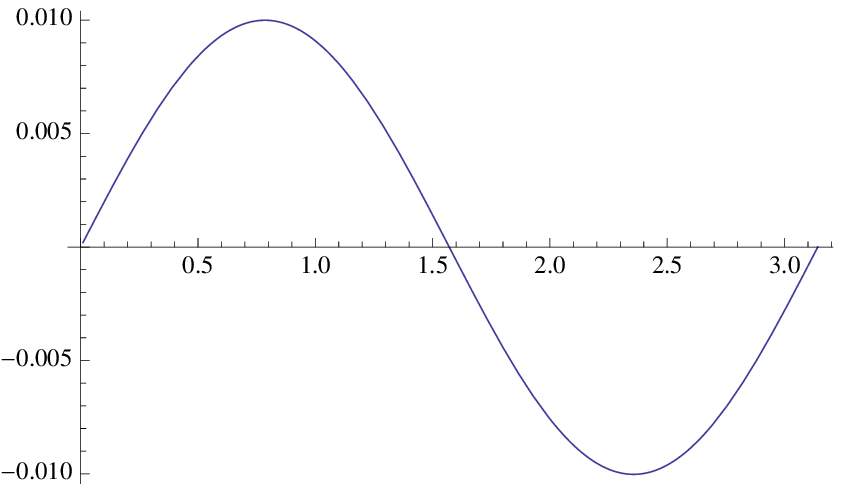}
\end{minipage}%
\begin{minipage}[t]{0.5\linewidth}
\vspace{0pt}
\centering
\includegraphics[scale=0.8]{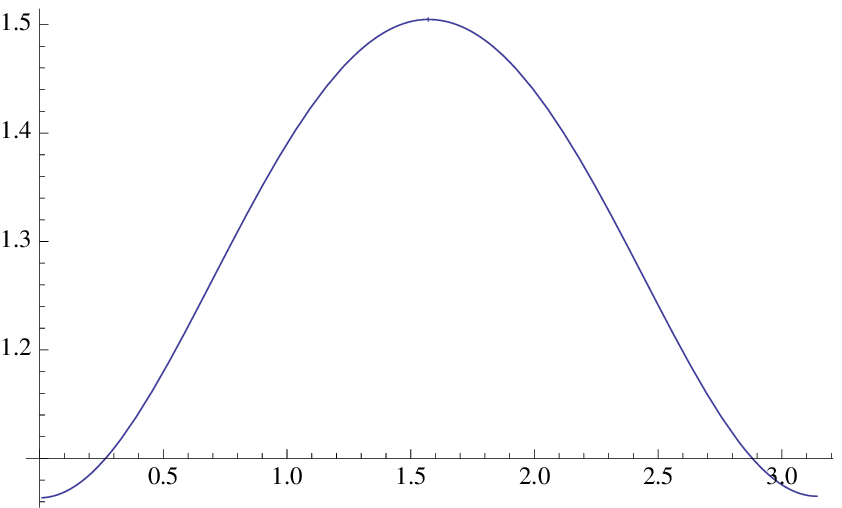}
\end{minipage}\\[1em]
\begin{minipage}[t]{0.5\linewidth}
\vspace{0pt}
\centering
\end{minipage}%

\vspace{1em}
Figure 1. Graphics of $\phi$ and $e^{2 A}$ for the values of parameters $\phi_1= 0.02$ and $w_0=4$.
The symmetry $\theta\rightarrow \pi - \theta$ is only present for the value $w_0= 4$.
\end{minipage}
\end{figure}

The solution depends on four arbitrary parameters: the asymptotic value of the dilaton, $g_s$, and
the radii of $AdS_4$, $\mathbb{P}^2$ and $S^2$ (these are functions of $f_0,\phi_1,w_0$). To these, we can add the zero modes 
of the $B$-field on the two two-cycles in the solution\footnote{We thank Alessandro Tomasiello for an enlightening discussion on this point.}, for a total of six parameters. Their values are constrained by flux quantization. $f_0$ is interpreted as  the period of a zero form 
and must be an integer. For the other RR forms, we define Page charges associated to the quantities $\tilde F_2=F_2-B f_0$, 
$\tilde F_4=F_4-B\wedge F_2$ and $\tilde F_6 =F_6-B\wedge F_4$. As we already said, the Bianchi identities are satisfied and these forms are closed. It follows from the ansatz (\ref{flux2def}) that we can conveniently write them as
 \beq 
 F_2-B f_0 = \d \Big[ f_2 \frac{(\d \tau +\lambda)}{16}\Big] \, , \qquad\, \qquad 
F_4-B\wedge F_2 = \d \Big[ f_4 \frac{(\d\tau +\lambda)}{16}\Big] \, . \label{quant} 
\eeq
We have to impose that the periods of these forms on all non trivial internal cycles are integer (in suitable units). 
A basis for the two-cycles is given by a copy of $\mathbb{P}^1\in \mathbb{P}^2$  at $\theta=0$ and by a copy of $S^2$ at a chosen  point in $\mathbb{P}^2$. Similarly, a basis for the four-cycles is given by $\mathbb{P}^2$
and $\mathbb{P}^1\times S^2$. Equation (\ref{quant}) allows to evaluate easily the periods. For example, without including the zero modes of the $B$ field, we have
\beq 
\int_{S^2} \tilde F_2 =  [f_2(\pi)-f_2(0)] \frac{\pi}{4} \, ,\qquad\,  \qquad \int_{\mathbb{P}^1} \tilde F_2 = f_2(0) \int_{\mathbb{P}^1} j_0 \, .
\eeq
All these integrals can be expressed in terms of the parameters of the solutions. The the zero modes of $B$ can be easily included. 
We obtain six quantization conditions for the six parameters in the solution. \\

Now we can compare the results with the field theory expectations. The original Chern-Simons theory can be deformed preserving 
$\mathcal{N}=2$ supersymmetry 
by changing  the ranks $N_i$ and by relaxing the condition $\sum_a k_a=0$. Since the  superpotential is unchanged, the $SU(3)$ global 
symmetry is also preserved. The most general $SU(3)$ invariant $\mathcal{N}=2$ Chern-Simons quiver depends on six integer parameters, the three ranks $N_i$ of the gauge groups and the three Chern-Simons couplings $k_a$.   We have room to describe all these in our deformed 
solution. The precise identification of the supergravity parameters with the field theory ones is complicated. A very rough 
identification is as follows. The original parameters $N$ and $k$ are still described by the constants in  the dilaton and the $AdS$ radius. The $\sum_a k_a$ can be associated with $f_0$ \cite{Gaiotto:2009mv}, while the difference between the ranks of the gauge groups 
should correspond to  the zero modes of the $B$-field \cite{Aharony:2008gk}.  The extra parameter corresponds to 
varying the ratio of  the $\tilde F_2$ periods on the two-cycles. In the unperturbed solution, this would correspond to replacing
$M^{111}$ with a generic member of the family $Y^{p,q}(\mathbb{P}^2)$. It is reasonable that our solution for  $f_0\ne 0$ already describes the  quiver with generic $k_i$, as the  parameters of the supergravity solution suggest. 
In fact, one can explicitly verify that our set of equations in the limit $f_0\rightarrow 0$ contain, in addition to the solution given in Section 3 with $w=4$, other solutions with non-trivial $w$ corresponding to the dimensional reduction of the manifolds $Y^{p,q}(\mathbb{P}^2)$ for generic $(k_1,k_2,-k_1-k_2)$ \cite{Gauntlett:2004hh, Martelli:2008rt}.
The different values of $k_a$ appear in type IIA  as periods of $\tilde F_2$.

As a final check, we can insert D2-brane probes in the background \cite{Gaiotto:2009yz}. As discussed in \cite{Martucci:2005ht,Martucci:2006ij,Gaiotto:2009yz}, the supersymmetry conditions for a probe D2 requires 
\beq
\d(\re \Phi |_{(0)})\equiv \d(\tan\rho) =0 \, .
\eeq 
Using  the perturbative and numerical expansion of the solution it is easy to check that a supersymmetric locus for D2 probes corresponds to $\theta=\pi/2$. On this locus, the $\tau$ fibration over $\mathbb{P}^2$ reproduces $S^5/\mathbb{Z}_3$. The moduli space for a supersymmetric D2 is obtained by adding the radial direction in $AdS_4$  giving the cone over $S^5/\mathbb{Z}_3$, which, as a complex three-dimensional  manifold, is  $\mathbb{C}^3/\mathbb{Z}_3$. 
This result matches the  moduli space of Chern-Simons quivers for $\sum_a k_a\ne0$, which
becomes three-dimensional. This is simple to see from equations (\ref{DtermII}). The sum of the three equations gives $\sigma \sum_a k_a=0$,
 which implies $\sigma=0$. The equations (\ref{DtermII}) become standard D-term constraints  for all gauge groups. In particular, we mod also by $U(1)_3$, reducing the complex dimension of the moduli space of one unit. The moduli space is then the same as for 
the $3+1$ dimensional superconformal theory based on the same quiver, that is $\mathbb{C}^3/\mathbb{Z}_3$ \footnote{Recall that the case $\sum_a k_a=0$, corresponding to zero Roman mass $f_0$,  is special: the moduli space of the quiver is of complex dimension four \cite{Jafferis:2008qz,Martelli:2008si,Hanany:2008cd} and we can uplift the solution to  M theory.
For $\sum_a k_a\ne 0$, differences between models seem to disappear: as discussed in the text, the moduli space of all $Y^{p,q}(\mathbb{P}^2)$  quivers degenerates to $\mathbb{C}^3/\mathbb{Z}_3$ and this  is well captured by the type IIA geometry we found.}.

\section{Conclusions}

In this paper we have considered an $AdS_4$  solution of massive type IIA supergravity with $\mathcal{N}=2$ supersymmetry and $SU(3)\times U(1)^2$ 
global symmetry. Having non zero Roman mass and non trivial dilaton, this solution has $SU(3)\times SU(3)$ structure.

The $AdS_4\times CFT_3$ correspondence actually suggests the existence of many supersymmetric $AdS_4$ vacua still to be found. The method
in this paper can be applied to various Sasaki-Einstein metrics. The generalization to $Y^{p,q}(\mathbb{P}^2)$ is already contained in our deformed equations. More interesting would be
to find solutions for the family   $Y^{p,q}(\mathbb{P}^1\times \mathbb{P}^1)$, which includes $Q^{111}$. 
In this case, the global symmetry is typically reduced
to $SU(2)^2$ and the corresponding solution  can be harder to find. Even in the apparently simple case of ABJM, the solution at all order is still lacking.
The perturbative method of \cite{Gaiotto:2009yz} however always applies. It would be interesting to perform a perturbative expansion for
the Sasaki-Einstein 
manifolds where a dual quiver has been proposed. This would be an interesting check of the correctness of the proposal, which is still unclear since
many standard checks of the correspondence cannot be fully performed in 2+1 dimensions.

One can also study deformations with $\mathcal{N}=1$ supersymmetry. Many solutions are still predicted by the correspondence,
since the dual quiver  can be typically deformed to $\mathcal{N}=1$, usually loosing some global symmetry. Alternatively, one can forget
about the dual interpretation, and search for $\mathcal{N}=1$ $SU(3)\times SU(3)$ solutions with large global symmetry. As shown in this paper, although 
apparently complicated, the supersymmetry conditions can be sometimes simplified and reduced to a simple set of equations. In particular, it would be
interesting to see whether there exist other $\mathcal{N}=1$ solutions of the equations we wrote with $SU(3)$ global symmetry. 
This is left for future work. \\
 
 \noindent {\bf Note added: \,}  As noted in \cite{Lust:2009mb}, $\mathbb{P}^2$ can be replaced by any compact   
 K\"ahler-Einstein base ${\cal B}$ without affecting the conditions of supersymmetry and without modifying any of the formulae in this paper.  In fact, all the equations follow from the existence of a trio of forms $\lambda,j_0,\omega_0$ on the base satisfying $\d \lambda = 16 j_0\, ,\, \d \omega_0 = i \frac{\lambda}{2} \wedge \omega_0$, which is a distinctive feature of  all K\"ahler-Einstein manifolds. For any M theory background $Y^{p,q}({\cal B})$, the equations in this paper  define  an $AdS_4$  massive type IIA deformation with non zero Roman mass and with the same global symmetries of ${\cal B}$. This agrees with the expectation that the corresponding quiver can be deformed by changing
the Chern-Simons couplings and the ranks of the gauge groups without breaking the global symmetries.
This construction applies in particular to ${\cal B}=\mathbb{P}^1\times \mathbb{P}^1$ which can be used to describe $Q^{111}$ and its quotients.

\section*{Acknowledgments}
We would like to thank A. Tomasiello for interesting discussions.  M. P.  is supported in part by ANR grant BLAN05-0079-01. A.Z. would like to thank the
NYU Physics Department  for hospitality and support during part of this
work. A.~Z.~ is supported in part by INFN and MIUR under contract 2007-5ATT78-002.

\begin{appendix}

\section{Conventions for $\mathbb{P}^2$}
\label{appp1}
 In our conventions, the metric on  $\mathbb{P}^2$ is normalized as
\beq
{\rm d}s^2_{\mathbb{P}^2} = \frac{3}{4} \,  [{\rm d}\mu^2 + \frac{1}{4} \sin^2 \mu \, \cos^2 \m^2 
( {\rm d} \psi + \cos \ttheta {\rm d} \tilde \phi)^2
+ \frac{1}{4} \sin^2 \mu ( {\rm d} \ttheta^2  + \sin^2 \ttheta {\rm d} \tilde\phi^2)]
\eeq
and $R^{\mathbb{P}^2}_{ab}=\frac{9}{2} g^{\mathbb{P}^2}_{ab}$. We chose vierbein,
\bea
e^1&=&\frac{\sqrt3}{4}\,r\sin\mu\cos\mu
(\d \psi+\cos\ttheta \, \d \tilde \phi ) \, , \nonumber \\
e^2&=&\frac{\sqrt3}{2}\,r\, \d\mu \, , \nonumber \\
e^3&=&\frac{\sqrt3}{4}\,r\sin\mu 
( \sin\psi \, \d\theta-\cos\psi\sin\ttheta \, \d\tilde \phi )  \, ,\nonumber \\
e^4&=&\frac{\sqrt3}{4}\,r\sin\mu
(\cos\psi \, \d\ttheta+\sin\psi\sin\ttheta \, \d\tilde \phi ) \, ,
\eea
The complex structure of $\mathbb{P}^2$ is given by $z_1=e_1+i e_2, z_2=e_3+i e_4$. We define 
the K\"ahler form and the holomorphic two form as
\bea
&& j_0=\frac{i}{2}( \d z_1\wedge \d\bar z_1+ \d z_2\wedge \d\bar z_2 ) \, , \\ 
&& \omega_0= \d z_1\wedge \d z_2 \, .
\eea

It is also convenient to define a rescaled $\hat \omega_0 = e^{i \tau/2} \omega_0$. A straightforward computation
gives the following useful relations:
\bea
&& \d j_0 = 0  \, ,\nonumber\\
&& \d \re\hat \omega_0 = -\frac{\d \tau+\lambda}{2}\wedge \im \hat\omega_0 \, ,\nonumber\\
&& \d \im\hat \omega_0 = \frac{\d \tau+\lambda}{2}\wedge \re \hat\omega_0 \, ,\nonumber\\
&& \d(\d\tau+\lambda) =  16 j_0 \, ,
\eea
where the connection one-form  $\lambda  = - 3 \sin^2 \mu \, ( {\rm d} \psi + \cos \ttheta \, {\rm d} \tilde\phi)$ satisfies
$\d \lambda =16 j_0$.
Let us notice that $j_0$ and $\lambda$ are $SU(3)$ invariant while $\hat\omega_0$ is not.

\section{Supersymmetry equations and pure spinors}
\label{apppure}

In this paper we will be interested in solutions of type IIA supergravity corresponding 
to warped products  of $AdS_4$ with an  internal compact manifold  
\beq
\d s^2_{10} = e^{2 A}  \d s^2_4 +  \d s^2_{6} \, ,
\eeq
where $A$ is the warp factor.
  
The solutions are also characterised by  non-trivial values for some of fluxes, respecting the symmetries of $AdS_4$.
The NS $H$-field has only internal indices 
and the RR fields split
\beq
\label{RRsplit}
F_p^{(10)} = \mathrm{\rm vol}_4 \wedge \hat{F}_{p-4} + F_p  \, ,
\eeq
where  ${\rm vol}_4$ is the unwarped four-dimensional volume. We can use Hodge duality to express 
the RR fluxes in terms of the internal components only
\beq
\hat{F}_{p - 4}= \lambda(\ast_6 F_{6-p}) \, ,
\eeq
 where $\lambda$ acts on forms as  the reversal of all indices
$\lambda(F_p)=(-1)^{\mathrm{Int}[p/2]} F_p$. \\

Generically, a supersymmetric solution of type II supergravity can be characterised 
by the form of the spinorial
parameters solving the supersymmetry constraints. For backgrounds which are of product type, such parameters
factorise accordingly into 4 and 6-dimensional spinors. For type IIA a suitable ansatz is  
\beq
\label{spinan} 
\begin{array}{c}
\epsilon_1 =\zeta_+ \otimes \eta^1_+ +  \zeta_- \otimes \eta^1_- \, ,\\
\epsilon_2 = \zeta_- \otimes \eta^2_+ +  \zeta_+ \otimes \eta^2_-  \, ,
\end{array}
\eeq
where  $\zeta_+$ is a four dimensional Weyl spinor and $\eta^i_+$, with $i=1,2$, 
are two a priori independent six-dimensional Weyl spinors. The subscripts $\pm$ denote 
positive and negative chirality spinors, in four and six dimensions.

The spinors  $\eta^1$ and $\eta^2$ define an $SU(3)$ structure on $M$, each. The intersection of the two will define
an $SU(2)$ structure on $M$ and a vector $z$. We can write
\beq
\label{SU2strsp}
\begin{array}{c}
\eta_{1 \, +} = a \eta_+ + b \chi_+ \, ,\\
\eta_{2 \, +} = x \eta_+ + y \chi_+ \, ,
\end{array}
\eeq
where $ \chi_+ = \frac{1}{2} z \cdot \eta_-  $.
$\eta_-$ is the complex conjugate of $\eta_+$ and $z \cdot$ denotes 
the Clifford multiplication by the one-form $z_m \gamma^m$.

If the two spinors are everywhere parallel, 
\beq
\label{SU3strsp}
\eta^1_+ = a \eta_+ \, , \qquad \qquad  \eta^2_+ = x \eta_+ \, ,
\eeq 
with $a$ and $b$ complex functions, the two $SU(3)$ structures are identified and 
the manifold admits an $SU(3)$ structure.\\

A convenient formalism to study supersymmetric flux backgrounds of this type in type II theories
is provided by Generalised Complex Geometry  \cite{Hitchin, gualtieri}. 
Given a 6-dimension manifold $M$, one considers the sum of tangent and cotangent bundles,
$TM \oplus T^\ast M$,  and then construct the corresponding spinors. These
are $Spin(6,6)$ and have a representation in terms of polyforms on $M$: positive and negative chirality spinors
will correspond to even and odd forms, respectively,
\beq
\label{polyf}
\Phi_\pm  \in \Lambda^{even/odd}(T^\ast M) \, .
\eeq

As far as supersymmetry is concerned, we will be interested
in pure spinors, namely vacua of the Clifford algebra. These can be constructed as tensor products of the supersymmetry
parameters
 $\eta^1$ and $\eta^2$
\beq
\label{purespindef}
\Phi_\pm = \eta^1_+ \otimes \eta^{2 \, \dagger}_2 \, .
\eeq
The pair of pure spinors \eqref{purespindef} are also compatible (they have three common annihilators) and therefore
define an $SU(3) \times SU(3)$ structure on $TM \oplus T^\ast M$.
In a way the the two $SU(3)$ can be seen as corresponding to the two $SU(3)$ structures associated to the spinors
$\eta^1$ and $\eta^2$. Then, depending on the relation between them, the explicit form of the pure spinors will change. 

For $SU(3)$ structure, the pure spinors have a particularly simple form
\bea
\label{SU(3)ps}
\Phi_+ &=& \frac{a \bar{x}}{8} \, e^{-i J} \, , \nonumber\\
\Phi_- &=& - \frac{i a x}{8} \, \Omega  \, ,   
\eea
where $J$ and $\Omega$ are the K\"ahler form and the holomorphic three-form on the manifold.
For the general $SU(2)$ case, the pure spinors read
\bea
\label{purespSU2}
&& \Phi_+  = \frac{1}{8} \Big[ a \bar x e^{- i j} +  b \bar y e^{i j} - i
( a \bar{y} \omega + \bar{x} b \bar{\omega} ) \Big] e^{1/2 z \bar{z}} \, , \nonumber \\
&& \Phi_{-}  = \frac{1}{8} \Big[ i (b y \bar{\omega} - a x \omega) + ( b x  
e^{i j} - a y e^{- i j}) \Big] z\, ,
\eea
with $|a|^2+|b|^2=|x|^2+|y|^2=e^{A}$. \\

Supersymmetric vacua can be found by solving the ten-dimensional supersymmetry constraints and the Bianchi identities
for the NS and RR fluxes.
In \cite{gmpt1,gmpt2}, it was shown that the supersymmetry conditions are 
equivalent to a set of differential equations for the pure spinors $\Phi_\pm$ on the internal manifold $M$.
In type IIA such conditions read 
\bea
\label{susyeq1app}
&& ({\rm d}-H\wedge) (e^{A-\phi} \re\Phi_- ) = 0 \, ,\\
\label{susyeq2app}
&& ({\rm d}-H\wedge) (e^{3A-\varphi} \im \Phi_- ) = - 3 e^{2A - \varphi} \mu \im\Phi_+  + \frac{e^{4 A}}{8} \ast \lambda(F) \, ,\\
\label{susyeq3app}
&& ({\rm d}-H\wedge) (e^{2A-\varphi}  \Phi_+)= - 2  \mu  e^{A - \varphi}  \re\Phi_-  \, ,
\eea
where $\mu$ is related to the cosmological constant
$\Lambda$ as $\Lambda =- 3 |\mu|^2$.  Notice that the first equation is actually implied by the last
one.

Similarly the Bianchi identities can be given in terms of the internal fluxes only
\beq
\label{BI}
\d H=0  \qquad \qquad (\d - H\wedge) F = 0 \, ,
\eeq
where $F$ is the sum of all internal RR field strengths
\beq
F = F_0 + F_2 + F_4 + F_6 \, .
\eeq

\end{appendix}


\begin{thebibliography}{99}




\bibitem{Aharony:2008ug}
  O.~Aharony, O.~Bergman, D.~L.~Jafferis and J.~Maldacena,
  {\it N=6 superconformal Chern-Simons-matter theories, M2-branes and their
  gravity duals},
  arXiv:0806.1218 [hep-th].

\bibitem{Fabbri:1999hw}
  D.~Fabbri, P.~Fre', L.~Gualtieri, C.~Reina, A.~Tomasiello, A.~Zaffaroni and A.~Zampa,
  {\it 3D superconformal theories from Sasakian seven-manifolds: New  nontrivial
  evidences for AdS(4)/CFT(3)},
  Nucl.\ Phys.\  B {\bf 577} (2000) 547
  [arXiv:hep-th/9907219].


\bibitem{Ceresole:1999zg}
A. ~Ceresole, G. ~Dall'Agata, R. ~D'Auria and S. ~Ferrara,.
{\it M theory on the Stiefel manifold and 3-D conformal field theories},
JHEP  {\bf 0003} 011 (2000)
[arXiv:hep-th/9912107].

\bibitem{Billo:2000zr}
  M.~Billo, D.~Fabbri, P.~Fre, P.~Merlatti and A.~Zaffaroni,
  {\it Rings of short N = 3 superfields in three dimensions and M-theory on AdS(4) x N(0,1,0)},
  Class.\ Quant.\ Grav.\  {\bf 18} (2001) 1269
  [arXiv:hep-th/0005219].

\bibitem{Oh:1998qi}
K.~Oh and R.~Tatar,
``Three dimensional SCFT from M2 branes at conifold singularities,''
JHEP {\bf 9902}, 025 (1999)
[arXiv:hep-th/9810244].



\bibitem{Benna:2008zy}
  M.~Benna, I.~Klebanov, T.~Klose and M.~Smedback,
 {\it Superconformal Chern-Simons Theories and AdS$_4$/CFT$_3$ Correspondence},
 [arXiv:0806.1519 [hep-th]].


\bibitem{Imamura:2008nn}
  Y.~Imamura and K.~Kimura,
 {\it On the moduli space of elliptic Maxwell-Chern-Simons theories},
 [arXiv:0806.3727 [hep-th]].


\bibitem{Hosomichi:2008jd}
 K.~Hosomichi, K.~M.~Lee, S.~Lee, S.~Lee and J.~Park,
 {\it N=4 Superconformal Chern-Simons Theories with Hyper and Twisted Hyper Multiplets},
JHEP {\bf 0807} 091 (2008)
[arXiv:0805.3662 [hep-th]]. 


\bibitem{Hosomichi:2008jb}
  K.~Hosomichi, K.~M.~Lee, S.~Lee, S.~Lee and J.~Park,
 {\it N=5,6 Superconformal Chern-Simons Theories and M2-branes on Orbifolds},
 [arXiv:0806.4977 [hep-th]].

\bibitem{Terashima:2008ba}
  S.~Terashima and F.~Yagi,
{\it Orbifolding the Membrane Action},
 [arXiv:0807.0368 [hep-th]].

\bibitem{Hanany:2008qc}
  A.~Hanany, N.~Mekareeya and A.~Zaffaroni,
 {\it Partition Functions for Membrane Theories},
  JHEP {\bf 0809}, 090 (2008)
  [arXiv:0806.4212 [hep-th]].


\bibitem{Aharony:2008gk}
  O.~Aharony, O.~Bergman and D.~L.~Jafferis,
 {\it Fractional M2-branes},
 [arXiv:0807.4924 [hep-th]].

\bibitem{Jafferis:2008qz}
  D.~L.~Jafferis and A.~Tomasiello,
 {\it A simple class of N=3 gauge/gravity duals},
  JHEP {\bf 0810}, 101 (2008)
  [arXiv:0808.0864 [hep-th]].


\bibitem{Fuji:2008yj}
H. ~Fuji, S. ~Terashima  and  M. ~Yamazaki,
{\it A New N=4 Membrane Action via Orbifold},
Nucl. \ Phys. \ B {\bf 810}, 354 (2009)
[arXiv:0805.1997 [hep-th]]. 



\bibitem{Martelli:2008si}
  D.~Martelli and J.~Sparks,
{\it Moduli spaces of Chern-Simons quiver gauge theories and AdS(4)/CFT(3)},
  Phys.\ Rev.\  D {\bf 78}, 126005 (2008)
  [arXiv:0808.0912 [hep-th]].



\bibitem{Hanany:2008cd}
  A.~Hanany and A.~Zaffaroni,
{\it Tilings, Chern-Simons Theories and M2 Branes},
  JHEP {\bf 0810}, 111 (2008)
  [arXiv:0808.1244 [hep-th]].



\bibitem{Hanany:2008fj}
  A.~Hanany, D.~Vegh and A.~Zaffaroni,
{\it Brane Tilings and M2 Branes},
  JHEP {\bf 0903}, 012 (2009)
  [arXiv:0809.1440 [hep-th]].




\bibitem{Franco:2008um}
  S.~Franco, A.~Hanany, J.~Park and D.~Rodriguez-Gomez,
{\it Towards M2-brane Theories for Generic Toric Singularities},
  JHEP {\bf 0812}, 110 (2008)
  [arXiv:0809.3237 [hep-th]].




\bibitem{Hanany:2008gx}
  A.~Hanany and Y.~H.~He,
 "M2-Branes and Quiver Chern-Simons: A Taxonomic Study,''
  arXiv:0811.4044 [hep-th].

\bibitem{Davey:2009sr}
  J.~Davey, A.~Hanany, N.~Mekareeya and G.~Torri,
  "Phases of M2-brane Theories,''
  arXiv:0903.3234 [hep-th].


\bibitem{Franco:2009sp}
  S.~Franco, I.~R.~Klebanov and D.~Rodriguez-Gomez,
{\it M2-branes on Orbifolds of the Cone over Q$^{1,1,1}$},
 [arXiv:0903.3231 [hep-th]].

\bibitem{Amariti:2009rb}
  A.~Amariti, D.~Forcella, L.~Girardello and A.~Mariotti,
  arXiv:0903.3222 [hep-th].


\bibitem{BC04}
 K. ~Behrndt and M. ~Cvetic,
{\it General N=1 supersymmetric fluxes in massive type IIA string theory},
Nucl. \ Phys. \  B {\bf 708} 45 (2005)
[arXiv:hep-th/0407263].


\bibitem{LT04} 
D. ~L\"ust and D. Tsimpis,
{\it Supersymmetric AdS(4) compactifications of IIA supergravity},
JHEP {\bf 0502} 027 (2005)
[arXiv:hep-th/0412250].



\bibitem{Tomasiello:2007eq}
  A.~Tomasiello,
{\it New string vacua from twistor spaces},
  Phys.\ Rev.\  D {\bf 78}, 046007 (2008)
  [arXiv:0712.1396 [hep-th]].


\bibitem{KLT08}
P. ~Koerber, D. ~L\"ust and D. Tsimpis,
{\it Type IIA AdS(4) compactifications on cosets, interpolations and domain walls},
JHEP {\bf 0807} 017 (2008)
[sarXiv:0804.0614 [hep-th]]. 

\bibitem{Koerber09}
P. ~Koerber,
{\it A new family of non-homogeneous type IIA flux vacua from AdS4/CFT3}
[arXiv:0904.0012 [hep-th]].


\bibitem{LT09} 
D. ~L\"ust and D. Tsimpis,
{\it  Classes of AdS(4) type IIA/IIB compactifications with SU(3) x SU(3) structure}
[arXiv:0901.4474 [hep-th]].


\bibitem{Gaiotto:2009mv}
  D.~Gaiotto and A.~Tomasiello,
 {\it The gauge dual of Romans mass},
  arXiv:0901.0969 [hep-th].

\bibitem{Gaiotto:2009yz}
  D.~Gaiotto and A.~Tomasiello,
  ``Perturbing gauge/gravity duals by a Romans mass,''
  arXiv:0904.3959 [hep-th].

\bibitem{Gauntlett:2004hh}
  J.~P.~Gauntlett, D.~Martelli, J.~F.~Sparks and D.~Waldram,
  Adv.\ Theor.\ Math.\ Phys.\  {\bf 8}, 987 (2006)
  [arXiv:hep-th/0403038].


\bibitem{Martelli:2008rt}
  D.~Martelli and J.~Sparks,
  ``Notes on toric Sasaki-Einstein seven-manifolds and $AdS_4/CFT_3$,''
  JHEP {\bf 0811}, 016 (2008)
  [arXiv:0808.0904 [hep-th]].


 


\bibitem{Castellani:1983mf}
  L.~Castellani, R.~D'Auria and P.~Fre,
 {\it SU(3) X SU(2) X U(1) From D = 11 Supergravity},
  Nucl.\ Phys.\  B {\bf 239}, 610 (1984).

\bibitem{Page:1984ad}
  D.~N.~Page and C.~N.~Pope,
 {\it Stability Analysis Of Compactifications Of D = 11 Supergravity With SU(3) X SU(2) X U(1) Symmetry},
  Phys.\ Lett.\  B {\bf 145}, 337 (1984).
\bibitem{Forcella:2008bb}
  D.~Forcella, A.~Hanany, Y.~H.~He, and A.~Zaffaroni,
  ``The master space of $N=1$ gauge theories,''
  arXiv:0801.1585 [hep-th].



\bibitem{Nilsson:1984bj}
  B.~E.~W.~Nilsson and C.~N.~Pope,
  ``Hopf Fibration Of Eleven-Dimensional Supergravity,''
  Class.\ Quant.\ Grav.\  {\bf 1}, 499 (1984).

\bibitem{Sorokin:1985ap}
  D.~P.~Sorokin, V.~I.~Tkach and D.~V.~Volkov,
  ``On The Relationship Between Compactified Vacua Of D = 11 And D = 10
  Phys.\ Lett.\  B {\bf 161}, 301 (1985).





\bibitem{Hitchin}
N.~Hitchin,
{\it Generalized Calabi-Yau manifolds},
Quart. \ J.Math. \ Oxford Ser. {\bf 54},  281 (2003)
[arXiv:math.dg/0209099].

\bibitem{gualtieri}
M.~Gualtieri,
{\it Generalized complex geometry},
Oxford University DPhil thesis, 
[arXiv:math.DG/0401221].

\bibitem{gmpt1}
 M.~Grana, R.~Minasian, M.~Petrini and A.~Tomasiello,
 {\it  Supersymmetric backgrounds from generalized Calabi-Yau manifolds},
  JHEP {\bf 0408} (2004) 046,
  arXiv:hep-th/0406137.

\bibitem{gmpt2}
  M.~Grana, R.~Minasian, M.~Petrini and A.~Tomasiello,
 {\it Generalized structures of $\mathcal N = 1$ vacua},
  JHEP {\bf 0511} (2005) 020,
  arXiv:hep-th/0505212.

\bibitem{Minasian:2006hv}
  R.~Minasian, M.~Petrini and A.~Zaffaroni,
 {\it Gravity duals to deformed SYM theories and generalized complex geometry},
  JHEP {\bf 0612}, 055 (2006)
  [arXiv:hep-th/0606257].

\bibitem{Martucci:2005ht}
  L.~Martucci and P.~Smyth,
 {\it Supersymmetric D-branes and calibrations on general N = 1 backgrounds},
  JHEP {\bf 0511}, 048 (2005)
  [arXiv:hep-th/0507099].


\bibitem{Martucci:2006ij}
  L.~Martucci,
 {\it D-branes on general N = 1 backgrounds: Superpotentials and D-terms},
  JHEP {\bf 0606}, 033 (2006)
  [arXiv:hep-th/0602129].

\bibitem{Lust:2009mb}
  D.~Lust and D.~Tsimpis,
  {\it New supersymmetric AdS4 type II vacua},
  [arXiv:0906.2561 [hep-th]].

















\end{thebibliography}
\end{document}